\documentclass[apj,twocolumn]{emulateapj}
\usepackage{natbib}
\usepackage{amsmath}
\usepackage{multirow}

\usepackage{bm}
\usepackage{color}
\usepackage{dcolumn}
\usepackage{graphicx}
\usepackage{hyperref}

\newif\ifAMStwofonts
\AMStwofontstrue 
\def\vgrad{{\bm \nabla}}
\def\vgrad{{\bm \nabla}}

\def\cN{{\cal N}}

\def\vB{{\bf B}}

\def\gsim{~\rlap{$>$}{\lower 1.0ex\hbox{$\sim$}}}

\def\simpropto{\lower.2ex\hbox{$\; \buildrel \propto \over \sim \;$}}
\def\ltsim{\lower.5ex\hbox{$\; \buildrel < \over \sim \;$}}
\def\gtsim{\lower.5ex\hbox{$\; \buildrel > \over \sim \;$}}
\def\ltsim{\lower.5ex\hbox{$\; \buildrel < \over \sim \;$}}
\def\gtsim{\lower.5ex\hbox{$\; \buildrel > \over \sim \;$}}

\def\kms{\mbox{km\,s$^{-1}$}}

\def\dd{\,{\rm d}}

\def\kms{\ {\rm km\,s^{-1}}}
\def\hmpc{\ {\rm h^{-1}Mpc}}

\def\dd{{\rm d}}

\def\pmb#1{\setbox0=\hbox{#1}%
\kern-.025em\copy0\kern-\wd0
\kern.05em\copy0\kern-\wd0
\kern-.025em\raise.0433em\box0}

\def\vv{\pmb{$v$}}

\def\vv{\pmb{$v$}}

\def\vY{\pmb{$Y$}}
\def\vpsi{\pmb{$\Psi$}}
\def\vtheta{\pmb{$\theta$}}
\def\vvarphi{\pmb{$\varphi$}}

\def\vx{\pmb{$x$}}
\def\vy{\pmb{$y$}}
\def\vz{\pmb{$z$}}
\def\vr{\pmb{$r$}}

\def\vS{\pmb{$S$}}

\def\hvr{\hat {\vr}}

\def\hvx{\hat {\vx}}
\def\hvy{\hat {\vy}}
\def\hvz{\hat {\vz}}

\def\vpsi{\pmb{$\psi$}}

\def\simlt{\lower.5ex\hbox{$\; \buildrel < \over \sim \;$}}
\def\simgt{\lower.5ex\hbox{$\; \buildrel > \over \sim \;$}}

\newcommand{\beq}{\begin{equation}}
\newcommand{\eeq}{\end{equation}}
\def\beqa{\begin{eqnarray}}
\def\eeqa{\end{eqnarray}}
\def\fixit#1{}
\def\hmpc{h^{-1}\,{\rm Mpc}}

\def\dd{{\rm d}}

\shorttitle{Bulk flow estimation and mean motions }
\shortauthors{Nusser}
\begin{document}
\title{An Inconsistency in standard Maximum Likelihood Estimation of Bulk Flows }
\author{Adi Nusser }
\email{adi@physics.technion.ac.il}
\affil{Physics Department and the Asher Space Science Institute-Technion, Haifa 32000, Israel\\
e-mail: adi@physics.technion.ac.il}

\begin{abstract}
Maximum Likelihood estimation of the bulk flow  from radial peculiar motions of galaxies, 
generally assumes a constant velocity field inside the survey volume.  The assumption is inconsistent with 
the definition of  the bulk flow as the average of the  peculiar velocity field over the relevant volume.
This follows from a straightforward mathematical relation between the bulk flow of a sphere and the velocity potential on its surface.
The inconsistency exists also  for ideal data with exact radial velocities and 
full spatial coverage. 
 Based on the same relation we propose a simple modification to correct for this inconsistency. 
 \end{abstract}


\maketitle
\section{Introduction}
In the standard cosmological paradigm, galaxies share the same peculiar velocity field (deviations from a pure Hubble flow), $\vv$, 
as the underlying dark matter, at least on large scales away from virialized regions. 
Thus, although the distribution of galaxies may not be an honest tracer of the underlying density field, their motions 
offer, in principle, an unbiased probe of the gravitationally dominant dark matter.
 The bulk flow, defined as   the average  peculiar velocity in a volume of space, is one of the common statistical measures of 
  the velocity field.
  Usually, the bulk flow of a sphere of comoving radius
$r$ centered at the observer is considered,  
\begin{equation}
\label{eq:Bdef}
\vB(r)=\frac{3}{4\pi r^3}\int _0^r \vv(\vr') \dd^3 r'\; .
\end{equation}
The is the  most basic velocity moment beyond the trivial monopole  term describing a purely  radial flow.
Nonetheless, its estimation from observational data is a non-trivial matter.
The  relevant observations are very challenging and  provide only the radial (line of sight component)  peculiar motions 
of  a  relatively small number (a few $\sim 10^3$) of galaxies, within  $\ltsim 200\hmpc$ \citep[e.g][]{mas06,CF2}.  
Analyses of these observations  could easily be plagued by systematic biases 
due to  sparseness of  data and varying quality of distance measurements \citep{lyn88}.
The  bulk flow is essentially a large scale moment and hence it is particularly prone for 
systematics  which may masquerade as a real  signal. 
Putting these potential biases aside, we focus here on a single point related to estimating $\vB$ from 
velocity data by means of a Maximum Likelihood (ML) estimation \citep[e.g.][]{K88}. 
We address the issue in view of a very simple relation between $\vB(r)$ and the velocity potential 
on the surface of the sphere of radius $r$. 
Assuming a constant bulk flow in the sphere, the ML estimation 
does not yield the average flow as given in (\ref{eq:Bdef}).   However, we will see that there exists a simple remedy  for this 
inconsistency. 
\section{Basics}
\label{sec:basics}
Let $\hvx$, $\hvy$ and $ \hvz$ 
be unit vectors in the three axes of fixed Cartesian system. 
The radial direction  is indicated by the unit vector $\hvr$ and 
the projections of $\hvr$ into the cartesian axes are $\hat n^\alpha$, where $\alpha$ runs over $x,y$ and $z$, corresponding to 
$\hat n^x= \hvr\cdot \hvx =\sin \theta \cos \varphi$, $\hat n^y=\hvr\cdot \hvy= \sin \theta \sin \varphi$
and $\hat n^z=\hvr\cdot \hvz=\cos \theta $. These projections can be represented as combinations of the $l=1$ spherical harmonics, { $\hat n_z=\sqrt{4\pi/3}Y_{_{10}}$, $\hat n_y= i \sqrt{2\pi/3}(Y_{_{11}}+Y_{_{1\, -1}})$ and 
 $\hat n_x=- \sqrt{2\pi/3}(Y_{_{11}}-Y_{_{1\, -1}})$.  } 

We assume   a potential flow, i.e. the peculiar velocity can be written as  $\vv(\vr)=-\vgrad \phi(\vr)$, where $\phi$ is the 
velocity potential function. 
If we expand the angular dependence of $\phi$ into spherical harmonics,$Y_{lm}$, then it is easy to see that the only
term contributing to $\vB$ in (\ref{eq:Bphi}) is the dipole, $l=1$. 
The expansion by means of  $l=1$ spherical harmonics is entirely equivalent to 
a representation  in terms of the angular functions $n^\alpha(\theta,\varphi)$
\begin{equation}
\label{eq:phialpha}
\phi(r,\hvr)=\sum_{\alpha\in x,y,z}^3 \phi^\alpha(r)  \hat n^\alpha \; , 
\end{equation}
where 
\begin{equation}
\phi^\alpha(r)=\frac{3}{4\pi}\int\dd \Omega \phi(\vr) n^\alpha
\end{equation}
thanks to orthogonality conditions 
$
\int \dd \Omega n^\alpha n^\beta={4\pi}/{3} \delta^{\rm K}_{\alpha \beta}$.
The corresponding representation of the peculiar velocity field is
\begin{equation}
\label{eq:vecy}
\vv=-\sum_\alpha \left[\frac{\dd \phi^\alpha}{\dd r}\hat n^\alpha \hvr+\frac{\phi^\alpha}{r} \vgrad_{_{\theta,\varphi}} \hat n^\alpha\right]\; ,
\end{equation}
where  
$\vgrad_{_{\theta,\varphi}} \hat n^\alpha=\partial \hat n^\alpha/\partial \theta \hat \vtheta +(\sin\theta)^{-1}\partial \hat n^\alpha/\partial \varphi
\hat \vvarphi$ is perpendicular to $\hvr$.   {This representation  is equivalent to an expansion  of $\vv$ in terms of vector spherical harmonics  $\vY$ and $\vpsi$.}

This relation implies that  the radial velocity $u(\vr) =\vv \cdot \hvr$ and $\dd \phi/\dd r$ are related by 
\begin{equation}
\label{eq:ualpha}
\frac{\dd \phi^\alpha}{\dd r}=-\frac{3}{4\pi}\int u(\vr) \hat n^\alpha \dd \Omega\; .
\end{equation} 
As an example for the representation in terms $\hat n^\alpha$,  consider  a constant velocity field $\vv=B_0 \hvz$ in the z-direction. 
In this case,  $\phi=\phi^z \hat n^z$ where $\phi^z(r)=-B_0 r$. A  substitution of this potential in (\ref{eq:vecy})
gives $\vv=B_0 \cos \theta \hvr _- B_0 \sin \theta  \hat \vtheta$ which gives $B^x=B^y=0$ and $B^z=B_0$, as expected. 

Both terms  on the r.h.s in the  relation (\ref{eq:vecy}) contribute to $\vB$ in (\ref{eq:Bdef}). We could integrate the relation over a sphere in order to get the bulk in terms of $\phi^\alpha$. However,  
a much more elegant way of achieving the same thing is via
the divergence theorem, which gives \citep{NDB14}
\begin{equation}
\label{eq:Bphi}
\vB(r)=-\frac{3}{4\pi r^3}\int_S\phi(\vr) \dd \vS\; ,
\end{equation} 
where the  integration is over the surface of the sphere of radius $r$, with surface element 
$\dd \vS=\dd \Omega r^2 \hvr$.
Substituting (\ref{eq:phialpha}) for $\phi(\vr)$,  this   relation   gives
\begin{equation}
\label{eq:Balpha}
B^\alpha=-\frac{\phi^\alpha}{r} \; . 
\end{equation}
where $B^\alpha$ correspond to the three Cartesian components $B_x$, $B_y$ and $B_z$.
The relation (\ref{eq:Bphi}) also gives the bulk flow of a thin spherical shell of radius $r$ as
\begin{equation}
\label{eq:Bshalpha}
B_{\rm sh}^\alpha=-\left[\frac{\dd \phi^\alpha}{\dd r} +2\frac{\phi^\alpha}{r}\right] \; . 
\end{equation}

\section{Bulk Flows from ML}
\label{sec:mle}
Assume we are provided with  galaxy positions $\vr_i$ and 
radial peculiar motions $u_i$ of $i=1\cdots N$ galaxies. The $1\sigma$ error on $u_i$ is $\sigma_i$ and 
we assume that the  positions are  given accurately. The latter assumption can be justified if we use the 
redshifts as proxy to $r_i$ rather than the observed distances.  ML provides  an estimate,  $\tilde \vB$,  of the bulk flow   by minimizing 
\begin{equation}
\label{eq:chisq}
\chi^2=\sum_i \frac{w_i}{\sigma_i^{2}} \left[ u_i - \tilde \vB\cdot \hvr_i\right]^2\; ,
\end{equation}
where we allow for a weighting the galaxies by $w_i$ in addition to the usual statistical weights 
dictated by $\sigma_i$.
At the minimum,  $\partial \chi^2/\partial \tilde B^\alpha=0 $ yields
  \begin{equation}
  \label{eq:chimin}
  \sum_\beta \sum_i w_i \sigma_i^{-2} \hat n^\alpha_i \hat n^\beta_i \tilde B^\beta=\sum_i w_i \sigma_i^{-2} u_i \hat n^\alpha_i
  \end{equation}
where we have used $\tilde \vB \cdot \hvr_i=\sum_\beta \tilde B^\beta \hat n^\beta$.
Observational errors typically depend on distance and not the angular position, hence $\sigma_i=
\sigma(r_i)$.  If  further  no angular selection is imposed on the observed galaxies and  $w_i=w(r_i)$,   then 
 the continuous limit  of (\ref{eq:chimin}) is
\begin{equation}
\label{eq:Bmlec}
\tilde B^\alpha=\frac{3\int \dd r' r'^2 \dd\Omega  \frac{w(r') \cN(r')}{\sigma^2(r')} u(\vr') \hat n^\alpha(\vr')}{4\pi\int \dd r' r'^2 \dd\Omega \frac{w(r) \cN(r')}{\sigma^2(r)}}\; , 
\end{equation}
where  $\cN(r)$, is the 3D number density of observed galaxies. We ignore here contribution of 
the underlying clustering of matter  and thus
 the dependence of $\cN$ on $r$ is entirely 
due to observational selection strategy. 

\section{The inconsistency and its resolution}
According to (\ref{eq:Bmlec}),  if   $u(\vr)=\vB_0\cdot \hvr$ where $\vB_0$ is constant throughout the sphere, then we recover $\tilde \vB=\vB_0$. 
However, the estimate   in (\ref{eq:Bmlec}) does not generally agree  with the definition of the bulk flow as given  in (\ref{eq:Bdef}).
To see this we rewrite   (\ref{eq:Balpha})  as  $ B^\alpha=-\phi^\alpha/r=-(\int  \dd r'  \dd \phi^\alpha/\dd r')/r$ and note 
the relation between $u$ and $\dd \phi^\alpha/\dd r$ in (\ref{eq:ualpha}).  
Hence, for a general choice of $w$, the mathematical relation   (\ref{eq:Balpha}) and the ML estimate  (\ref{eq:Bmlec}) are inconsistent.
 Nonetheless, the two equations become consistent for the specific 
  choice 
\begin{equation}
\label{eq:w}                        
w=\frac{\sigma^2}{\cN r^2}\; .
\end{equation}
 The term $\cN r^2$ could  be identified with the number density per unit radius. 
 Since typically  $\sigma \propto r$, the weighting is essentially  equivalent to $w\sim 1/\cN$. 
  Minimizing $\chi^2$ in (\ref{eq:chisq}) with the weights, $w$,  given by  (\ref{eq:w}) will yield    $\tilde \vB(r)$ that  is consistent   (up-to statistical error)  with the definition of the bulk flow as the average  
  peculiar velocity within the sphere.  In the absence of errors, this choice of $w$ guarantees that  $\tilde \vB(r)$ coincides with the true $\vB(r)$. 
  
 Another way to achieve an estimate  which agrees with the definition  
 (\ref{eq:Bdef}) is  as follows. 
 Let us divide space into a finite number, $s=1\cdots N_s$, of  spherical shells  each of radius $r_s$ and  thickness, $\delta_s\ll r_s$.
Define a new  $\chi^2$ function 
\begin{equation}
\chi^2\equiv \sum_{s=1}^{N_s}  \sum_{i\in s}\sigma_i^{-2}\left[u_i -\sum_{\alpha} V^{\alpha}_s \hat n^\alpha  \right]^2\; .
\end{equation}
Here the symbol $i\in s$ implies galaxies lying inside the shell $s$ and $V_s^{\alpha}=V^{\alpha}(r_s)$, where 
$V^\alpha=-\dd \phi^\alpha/\dd r$.
Minimization of this $\chi^2$ with respect to $V_s^\alpha$  gives 
\begin{equation}
\sum_{\beta} \sum_{i\in s}\sigma_i^{-2}  \hat n^\alpha_i   \hat n^\beta_iV_s^\beta=\sum_{i\in s}\sigma_i^{-2}u_i \hat n^\alpha_i\; .
\end{equation}
Once $V^\alpha_s$ are obtained by solving the last equations, the potential 
can be computed as $\phi^\alpha_s=-\sum \delta r_s  V^\alpha_s$ and
the bulk flow of a sphere of radius $r_s$ identified as $B^\alpha=-\phi^\alpha_s/r_s$.
Note that $V^\alpha_s$ coincides with the bulk flow of the shell only 
if the velocity in the shell is constant, otherwise, it will be missing the  term $-2\phi^\alpha/r$ as is seen from 
(\ref{eq:Bshalpha}). 

\section{A numerical demonstration} 
We give  a demonstration for the case of perfect data with zero errors and uniform spatial coverage.
We do that with the help of  a random gaussian realization of a 
velocity field with a power spectrum of the $\Lambda$CDM model with density parameters 
$ \Omega_c=0.225$, $\Omega_b=0.045$, and $\Omega_v=0.73$, respectively,  for the dark  matter, baryons and 
the cosmological constant. The field is generated on a $512^3$ uniform gris in a box of  $500\hmpc$ on the side, with $H_0=70\kms \rm  Mpc^{-1}$.  
 A grid point with a velocity close to the observed motion of the Local Group is chosen as the central ``observer".  
The 3D velocities at the grid points within a distance of $100\hmpc$ from the observer are used to directly compute the true bulk flow, 
$\vB_{\rm t}$, of spheres centered on the observer.
The actual radial velocities   at the  grid points are used as ``observational" data, without any dilution and 
and  any  added noise. Thus, $\cN(r)=const$ and $\sigma_i$ is a constant which formally is taken as very close to 
zero.
We then derive two  estimates for 
 bulk flows for spheres around the observer. The first estimate  is derived using  the standard ML as appropriate for this data  (i.e. $w\cN/\sigma^2=const$  in eq. \ref{eq:Bmlec}) 
and the second is obtained with the modified weighting in (\ref{eq:w}) (i.e. $w\cN/\sigma^2=1/r^2$ in  eq. \ref{eq:Bmlec}). 
The  two estimates and the true bulk flow are shown in the Figure as a function of the radius. 
The discrepancy between the standard ML estimate and the true bulk flow is  substantial, while the modified weighting 
almost yields perfect agreement. 
It is interesting that  the two estimates would coincide had the data been diluted to $\cN\propto r^{-2}$.

\begin{figure}
\includegraphics[width=.5\textwidth]{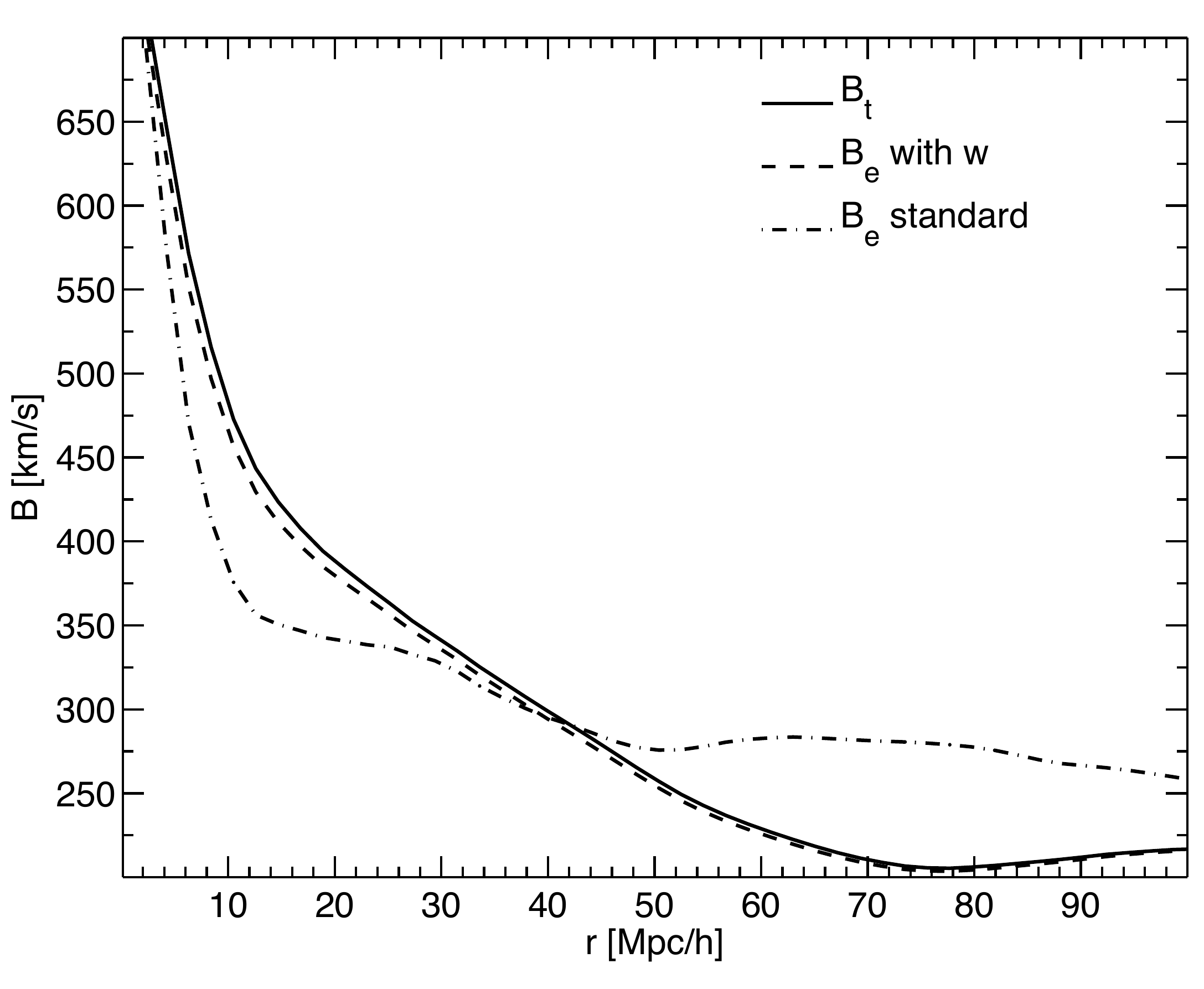}
\caption{Estimation of the bulk flow for ideal data taken from a realization of a random gaussian velocity field. 
The  bulk flow estimated with the weighting $w$ given by (\ref{eq:w}) agrees well with the true flow $\vB_t$. }
\end{figure}
\section{General  Remarks}
The inconsistency in the ML estimation pointed out here stems from the assumption of a constant $\vB$ in the survey volume. 
With this assumption, standard ML estimation yields, by definition,  the most likely {\it constant} velocity vector which fits the 
data inside the survey volume. However, this constant velocity  does not coincide with the definition of the bulk flow as the 
mean velocity of the relevant volume. 

We do not aim here at quantifying the inconsistency for realistic data. Datasets are available with numerous version, each with its own peculiar characteristics and the differences between results of various weightings should be assessed individually. 
Further, any weighting scheme could be applied to a given  dataset  as long 
as  the implications are  assessed self-consistently within the context of a cosmological model or in comparison with  other datasets. 
 However, to avoid confusion the term ``bulk flow" should be reserved to 
estimates of the mean motion rather than any other moment of the data. 

Peculiar velocity data  could be analyzed in many ways \citep[e.g.][]{DN10} which do not resort to an application of the ML estimation
as presented above. The constrained realizations method \citep[e.g][]{hr91,2014NewAR..58....1Y}  reconstructs  a full 3D velocity field from 
observed radial velocity data. In this method 
the bulk flow can be computed directly from the reconstructed 3D velocity  field.

\section{acknowledgment}
This research was supported by the I-CORE Program of the Planning and Budgeting Committee,
THE ISRAEL SCIENCE FOUNDATION (grants No. 1829/12 and No. 654/13)  and the Asher Space Research
Institute. The author thanks Enzo Branchini and Martin Feix for comments which helped improving the manuscript.
\bibliography{B}

\end{document}